\documentstyle{article}
\begin{document}
\tolerance 10000

\title{Kaluza-Klein 5D Ideas Made Fully Geometric}

\author{Scott A. Starks, Olga Kosheleva, 
and Vladik Kreinovich\\
NASA Pan-American Center for\\
Earth and Environmental Studies (PACES)\\ 
University of Texas at El Paso, El Paso, TX 79968, USA\\
\{sstarks,olgak,vladik\}@utep.edu}
\date{}
\maketitle

\begin{abstract}
After the 1916 success of General relativity that explained gravity by 
adding time as a fourth dimension, physicists have been trying to
explain other physical fields by adding extra dimensions. In 1921,
Kaluza and Klein has shown that under certain conditions like
cylindricity ($\partial g_{ij}/\partial x^5=0$), the addition of the
5th dimension can explain the electromagnetic field. The problem with
this approach is that while the model itself is geometric, conditions
like cylindricity are not geometric. This problem was partly solved by
Einstein and Bergman who proposed, in their 1938 paper, that the 5th
dimension is compactified into a small circle $S^1$ 
so that in the resulting 
cylindric 5D space-time $R^4\times S^1$ 
the dependence on $x^5$ is not macroscopically
noticeable. We show that if, in all definitions of vectors, tensors,
etc., we replace $R^4$ with $R^4\times S^1$, then conditions like
cylindricity automatically follow -- i.e., these conditions 
become fully geometric.
\end{abstract}

\section{Physics: 5D Geometry is Useful} 
After the 1916 success
of A. Einstein, who explained gravitation by combining space and time into
a 4D space, there have been many efforts to explain other physical
fields by adding other physical dimensions. 

The first successful
attempt was made by Th. Kaluza and O.~Klein in 1921. They showed that if we
formally consider the equations of general relativity theory in the 5D
space, the equations for the normal $4\times 4$ components
$g_{ij}$ of the
metric tensor still describe gravitation, while the new components
$g^{5i}$ of the metric tensor satisfy Maxwell's equations
(under the assumption that $g_{55}={\rm const}$). Thus, if
we go to 5D space, we get a geometric interpretation of electrodynamics. 

The only problem with this interpretation is that it is formal: change
in first 4 dimensions makes perfect physical sense, while there seemed
to be no physical effects corresponding to change in 5th dimension. To
solve this problem, A. Einstein and P. Bergmann 
proposed, in 1938 \cite{Einstein 1938}, that the 5th dimension
forms a tiny circle, so that only micro-particles ``see" it, while for
us, the world is 4D. 

This is a standard view now in particle physics; see, e.g.,
\cite{Green 1988,Polchinski 1998}: 
space is 10- or 11-dimensional, all dimensions except the first four
are tiny.

\section{Formulas from Physical 5D Theories that Need to Be
Explained in Purely Geometric Terms} 

In addition to a nice geometric model, the traditional description of
Kaluza-Klein theory requires several additional physical formulas, 
formulas that look very artificial because they do not have a direct
geometric explanation. 

In this paper, we will show that, if we take the Einstein-Bergmann
model seriously, then these formulas can be derived -- and thus, they
are not additional and ad hoc.  

What are these formulas that do not directly follow from the
geometric model? 

First, the assumption $g_{55}={\rm const}$ is artificial. 

Second, since only four coordinates have a physical sense, the
distance $\Delta s^2=\sum\limits_{i=1}^5\sum\limits_{j=1}^5 
g_{ij}\cdot \Delta x_i\cdot\Delta x_j$ between the points 
$x$ and $x+\Delta x$ should only depend on the first 4 coordinates --
while in general, for a 5D metric, the terms $g_{55}\cdot (\Delta
x^5)^2$ and $g_{5i}\cdot \Delta x^5\cdot\Delta x^i$ create a
difficult-to-explain dependence on $\Delta x^5$. 

Third, we would like to explain the fact that the observed values of
physical fields do not depend on the fifth coordinate $x^5$, e.g.,
that $\partial g_{ij}/\partial x^5=0$ (this condition is called {\it
cylindricity}). 

Several other formulas came from the attempts to give the fifth
dimension a physical interpretation. Namely,  
in the 1940s, Yu. Rumer showed (see, e.g., \cite{Rumer 1956})
that if we interpret $x^5$ as action
$S=\int L\,dx\,dt$ (i.e., the quantity whose extrema define the
field's dynamics), then the fact that $x^5$ is defined on a circle is
consistent with the fact that in quantum physics (e.g., in its Feynman
integral formulation), action is used only as part of the expression
$\exp(iS/h)$, whose value is not changed if we add a constant 
$2\pi\cdot h$ to $S$. (For a H atom, this idea leads to the original 
Bohr's quantization rules.) 

Action is defined modulo arbitrary transformation $S\to S+f(x^i)$;
thus, the corresponding transformation $x^5\to x^5+f(x^i)$ should be
geometrically meaningful. Similar transformations stem from the
electrodynamic interpretation of $g_{5i}$ as $A_i$: gauge
transformations $A_i\to A_i-\partial f/\partial x_i$. 

\section{Natural Idea and Its Problems} 

The main difference between a standard
4D space and Einstein-Bergmann's 5D model is that we have a cylinder 
$K=R^4\times S^1$ ($K$ for Kaluza)
instead of a linear space. It is, therefore,
desirable to modify standard geometry by substituting $K$ instead of
$R^4$ into all definitions. 

The problem with this idea is that the
corresponding formalisms of differential geometry use the underlying
linear space structure, i.e., addition and multiplication
by a scalar. We still have addition in $K$, but multiplication is not
uniquely defined for angle-valued variables: we can always interpret
an angle as a real number modulo the circumference, but then, e.g., $0\sim
2\pi$ while $0.6\cdot 0\not\sim 0.6\cdot 2\pi$. 

\section{What We Suggest} 
We do need a real-number
representation of an angle variable. A more natural representation of
this variable is not as a single real number, but as a {\it set}
$\{\alpha+n\cdot 2\pi\}$ of all possible real numbers that correspond
to the given angle.

Similarly to interval and fuzzy 
arithmetic, we can naturally define element-wise
arithmetic
operations on such sets, e.g., $A+B=\{a+b\,|\,a\in A, b\in B\}$. 
We can then define tensors as linear
mappings that preserve the structure of such sets, and we can define a
{\it differentiable} tensor field as a field for which the set of all
possible values of the corresponding partial derivatives is also
consistent with the basic structure. 
\medskip

\noindent{\it Comment.} These results were first announced
in \cite{Kreinovich 2005,Kreinovich 1997,Starks 1998}. 

\section{Resulting Formalism: Idea}
In mathematical terms, the
resulting formalism is equivalent to the following: 
We start with the space $K$ which is not a vector space (only an Abelian
group). We reformulate standard definitions of vector and
tensor algebra and tensor analysis and apply them to $K$: 
$K$-vectors are defined as elements of $K$; $K$-covectors as elements of
the dual group, etc.
All physically motivated conditions turn out to
be natural consequences of this formalism.

\section{$K$-Vectors}

In the traditional 4-D space-time $R^4$, we can define a {\it vector} as
simply an element of $R^4$. In our case, instead of 4-D space-time
$R^4$, we have a 5-D space-time $K\stackrel{\rm def}{=}R^4\times S^1$,
in which $S^1$ is a circle of a small circumference $h>0$ 
-- i.e., equivalently, a real line in which two numbers differing by a 
multiple of $h$ describe the same point: $(x^1,\ldots,x^4,x^5)\sim
(x^1,\ldots,x^4,x^5+k\cdot h)$. Thus, it is natural to define {\rm
$K$-vectors} as simply elements of $K$: 
\medskip

\noindent{\bf Definition 1.}
{\rm A {\em $K$-vector} is an element of $K=R^4\times S^1$.}
\medskip

On the set of all vectors in $R^4$, there are two natural operations:
(commutative) addition $a+b$ and multiplication by a real number
$\lambda$: $a\to \lambda\cdot a$. Thus, this set is a linear space. 

In contrast, on the the set $K$ of all $K$-vectors we only
have addition, so the set of all $K$-vectors
is {\it not} a linear space, it is only an Abelian group. 

\section{$K$-Covectors}

In physics, an important algebraic object is a {\it covector}:
vectors describe the location $x$ of a particle, 
while the corresponding covector $p$ describes the energy and momentum
of the corresponding particle. Because of this physical importance, it 
is necessary to generalize the notion of covectors to the new space. 

We would like to provide a generalization that preserves the physical
meaning of the connection between vectors and covectors. 
The physical connection is probably best described in quantum
mechanics. In quantum mechanics, due to Heisenberg's uncertainty
principle $\Delta x\cdot \Delta p\ge \hbar$, 
if we know the exact location of a particle (i.e., if $\Delta x=0$), 
then we have no information about the momentum (i.e., $\Delta
p=\infty$), and vice versa, if we know the exact momentum ($\Delta
p=0$), then we have no information about the particle's location. In
other words, if we have a state with a definite momentum $p$, and we
then shift the coordinates by a vector $t$, i.e., replace $x$ by
$x+t$, the known state of the particle should not change. 

In quantum mechanics, a state of the particle is described by a
complex-valued function $\psi(x)$ called a {\it wave function}. 
The wave function itself is
not directly observable, what we observe are probabilities
$|\psi|^2$. So, if we multiply all the values of the wave-function
by a complex number $\varphi$ with $|\varphi|=1$ (i.e., by a number 
of the type $\exp({\rm i}\cdot\alpha)$, where ${\rm
i}=\sqrt{-1}$ and $\alpha$ is a real number), then all the probabilities 
remain the same -- i.e., from the physical viewpoint, we will have
exactly the same state. Thus, for every real number $\alpha$, the functions
$\psi(x)$ and $\exp({\rm i}\cdot\alpha)\cdot \psi(x)$ describe exactly
the same state. When we say that the state $\psi(x)$ does not
change after shift $x\to x+t$, we mean that the original function
$\psi(x)$ and the function $\psi(x+t)$ that describe the shifted state 
describe the same state -- i.e., $\psi(x+t)=\varphi(t)\cdot\psi(x)=
\exp({\rm i}\cdot \alpha(t))\cdot\psi(x)$ 
for some complex number $\varphi(t)$ or, equivalently, 
real number $\alpha(t)$ (which, generally speaking, depends
on the shift $t$). 

Since $\exp({\rm i}\cdot 2\cdot \pi)=1$,
the value $\alpha(t)$ is only determined modulo $2\cdot \pi$. 
Thus, $\alpha(t)$ is a point on a
circle rather than a real number. 

For $x=0$, we get $\psi(t)=\varphi(t)\cdot\psi(0)$, so modulo a
multiplicative constant, shift-invariant states $\psi(t)$ 
are equal to the corresponding functions 
$\varphi(t)$. So, to determine such states, we must describe
all the corresponding functions $\varphi(t)$. 

When we shift by $t=0$, the function remains unchanged, i.e.,
$\varphi(0)=1$ (equivalently, $\alpha(0)=0$). 

If we first shift $t$ and then by $s$, then we get the same result as 
if we shift once by $t+s$. Hence, we have 
$$\varphi(s)\cdot (\varphi(t)\cdot 
\psi(x))=\varphi(t+s)\cdot \psi(x),$$ so
$\varphi(t+s)=\varphi(t)\cdot \varphi(s)$. So, from the physical
viewpoint, a shift-invariant state $\varphi$ 
is a mapping from $R^4$ to the unit 
circle $S^1=\{\varphi:|\varphi|=1\}$ that transform 0 into 1 and sum
into sum. In mathematics, such a mapping is called a {\it
homomorphism} from an Abelian additive group $R^4$ to $S^1$. 

It is also physically reasonable to assume that the wave
function is continuous -- hence, that the homomorphism $\varphi$ is
continuous. Continuous homomorphisms from an Abelian group $G$ to 
a unit circle are called {\it characters}; the set of all such
characters is also an Abelian group called {\it dual} (and denoted by
$G^*$). So, it is
natural to associate covectors with elements of the dual group. 

For $R^4$, this definition fits well with the more traditional one,
because 
it is known that for $R^4$, the dual group is also $R^4$: 
every character has the form $\exp({\rm i}\cdot p\cdot x)$. For
$K=R^4\times S^1$, we get a new definition: 
\medskip

\noindent{\bf Definition 2.} 
{\em A {\em $K$-covector} is a character of the group $K$, i.e., a
continuous homomorphism from $K$ to $S^1$. By a {\em sum} of two
covectors we mean the product of the corresponding
homomorphisms.}
\medskip

The set of all $K$-covectors is thus a dual group $K^*$ to $K$. It is
known that elements of this dual group have the form $\exp({\rm
i}\cdot p\cdot x)$, where $p=(p_1,\ldots,p_4,p_5)$, $p_1,\ldots,p_4$ 
can be
any real numbers, and $p_5$ is an multiple of $1/h$. Thus, the group
$K^*$ of all $K$-covectors is isomorphic to $R^4\times Z$, where $Z$
is the additive group of all integers.
\medskip

\noindent{\it Comment.} $K$-vectors
are simply elements $x=(x_1,\ldots,x_5)$ of $R^5$, some of which are
equivalent to each other: $x\sim x'$ if $x_5-x'_5=k\cdot h$ for some
integer $k$. In other words, a $K$-vector can be viewed as a set
$$\{x':x'\sim x\}=\{(x_1,\ldots,x_4,x_5+k\cdot h\}.$$ 

A unit circle $S^1$ can also be described as simply the set $R$ of all 
real numbers with the equivalence relation $\alpha\sim \alpha'$ 
if and only if $\alpha-\alpha'=k\cdot (2\cdot \pi)$ -- or,
equivalently, as the class of sets $\{\alpha+k\cdot (2\cdot\pi)\}$. 

In these terms, we can
alternative describe $K$-covectors as linear mappings
$x=(x_1,\ldots,x_5)\to p\cdot x=\sum p_i\cdot x_i$ 
from $R^5$ to $R$ that are consistent with the above structures, i.e., 
mapping for which $x\sim x'$ implies $p\cdot x\sim p\cdot x'$. 

\section{$K$-Tensors}

To describe individual particles, it is usually sufficient to consider 
vectors (that describe their location) and covectors (that describe
their momentum). However, to describe field 
theories such as Maxwell's theory of
electromagnetism or Einstein's General Relativity theory, it is not
sufficient to consider only vectors and covectors, we also need to
consider {\it tensors}. 

Specifically, for $G=R^4$, 
for every two integers $p\ge 0$ and $q\ge 0$, a tensor of 
valence $(p,q)$ can be defined as a {\it multi-linear} map $G^p\times
(G^*)^q\to R$ -- where {\it multi-linear} means that if we fix the
values of all the variables but one, we get a linear mapping. Every
such multi-linear mapping has the form 
$$x^{i_1},\ldots, y^{i_p},z_{j_1},\ldots,u_{j_q}\to $$
$$\sum_{i_1,\ldots,i_p,j_1,\ldots,j_q} t_{i_1\ldots i_p}^{j_1\ldots j_q} 
\cdot x^{i_1}\cdot \ldots \cdot y^{i_p}\cdot z_{j_1}\cdot \ldots\cdot
u_{j_q}$$
for some components $t_{i_1\ldots i_p}^{j_1\ldots j_q}$. 
We thus naturally arrive at the following definition:
\medskip

\noindent{\bf Definition 3.} {\em Let $G_1,\ldots,G_m,G$ be continuous 
Abelian groups. A mapping $t:G_1\times \ldots\times G_m\to G$ is
called {\em $Z$-multilinear} if for every $i$, if we fix the values of
all the variables except $i$-th, we get a homomorphism.}
\medskip

\noindent{\bf Definition 4.} {\em Let $p\ge 0$ and $q\ge 0$. 
By a {\em $K$-tensor} of valence $(p,q)$, we mean a continuous 
$Z$-multilinear mapping $t:K^p\times (K^*)^q\to S^1$.}
\medskip

\noindent{\it Comments.} For $R^4$ instead of $K$, 
this definition coincides with
the traditional one. 

When $K=R^4\times S^1$, this definition is consistent with the
previous ones: $K$-tensors of valence $(0,1)$ are $K$-covectors, and
$K$-tensors of valence $(1,0)$ are $K$-vectors. 

This definition can be reformulated as follows: a $K$-tensor is a
multi-linear mapping that is consistent with the equivalence sets 
structure, i.e., for
which $x\sim x',\ldots,y\sim y'$ implies that
$t(x,\ldots,y,z,\ldots,u)\sim t(x',\ldots,y',z,\ldots,u).$ 

Two multi-linear mappings $t$ and $t'$ describe the same $K$-tensor if 
$t(x,\ldots,y,z,\ldots,u)\sim t'(x,\ldots,y,z,\ldots,u)$ for all 
$x,\ldots,y,z,\ldots,u$.  
 
The following result describes all such mappings:
\medskip
\newpage

\noindent{\bf Proposition 1.} {\em 
\begin{itemize}
\item Every $K$-tensor has the form 
\end{itemize}
$$\exp\left({\rm i}\cdot 
\sum_{i_1,\ldots,i_p,j_1,\ldots,j_q} t_{i_1\ldots i_p}^{j_1\ldots j_q} 
\cdot x^{i_1}\cdot \ldots \cdot y^{i_p}\cdot z_{j_1}\cdot \ldots\cdot
u_{j_q}\right)$$
\begin{itemize}
\item[]for some components $t^{\ldots}_{\ldots}$. In this representation, 
of all the components in which
one of the lower indices is 5, only a component $t^{5\ldots 5}_5$ can
be non-zero, and it can only take values $2\cdot\pi\cdot h^{q-1}\cdot
k$ for some integer $k$. 
\item Vice versa, if we have a set of components $t^{\ldots}_{\ldots}$ in
which of all the components in which
one of the lower indices is 5, only a component $t^{5\ldots 5}_5$ 
may be non-zero, its value is $2\cdot\pi\cdot h^{q-1}\cdot
k$ for some integer $k$, then the above formula defines a $K$-tensor. 
\item Two sets of components 
$t^{\ldots}_{\ldots}$ and $s^{\ldots}_{\ldots}$
define the same $K$-tensor if and only if all their components
coincides with a possible exception of components $t^{5\ldots 5}$ and 
$s^{5\ldots 5}$ which may differ by $2\cdot \pi\cdot h^q\cdot k$ for
an integer $k$. 
\end{itemize}}
\medskip

\noindent{\it Comment.} For readers' convenience, all the proofs are
given in the Appendix. 

\section{Explaining the Condition $g_{55}={\rm const}$ and the Fact 
that Metric Does Not Depend on $x^5$}

For $g_{ij}$, Proposition 1 implies that $g_{55}=g_{5i}=0$. Thus, the
above geometric formalism explains the first two physical assumptions
that we wanted to explain: that $g_{55}=0$ and that the distance 
$\Delta s^2=\sum\limits_{i=1}^5\sum\limits_{j=1}^5 
g_{ij}\cdot \Delta x_i\cdot\Delta x_j$ between the two points 
$x$ and $x+\Delta x$ only depends on their first 4 coordinates.

\section{Differential Formalism for $K$-Tensor Fields}

\noindent{\bf Definition 5.} {\em By a {\em $K$-tensor field}
$f^{\ldots}_{\ldots}(x)$ of valence
$(p,q)$, we mean a mapping that assigns, to every point $x\in K$, a
$K$-tensor $f^{\ldots}_{\ldots}(x)$ of this valence.} 
\medskip

Most physics is described in the language of differential
equations. It is known that for every tensor field 
$t_{i_1\ldots i_p}^{j_1\ldots j_q}$ 
of valence $(p,q)$, its gradient 
$\partial t_{i_1\ldots i_p}^{j_1\ldots j_q}/\partial x^m$ 
is also a tensor field -- of
valence $(p+1,q)$. This new field is called a {\it gradient} tensor
field. It is therefore natural to give the following
definition:
\medskip

\noindent{\bf Definition 6.} {\em We say that a $K$-tensor field of
valence
$(p,q)$ is {\em differentiable} if the corresponding component tensor
field is continuously 
differentiable, and its gradient field also defines a
$K$-tensor field.}
\medskip

In other words, to differentiate a $K$-tensor field, we form the
corresponding tensor field, differentiate it, and then interpret the
result as a $K$-tensor field of valence $(p+1,q)$. When is this
possible? The answer to this question is as follows:
\medskip

\noindent{\bf Proposition 2.} {\em The $K$-tensor field is
differentiable if and only if all its components 
$t^{\ldots}_{\ldots}$ do
not depend on $x^5$, with the possible exception of the component
$t^{5\ldots 5}$ which may have the form $2\cdot \pi\cdot h^{q-1}\cdot
x^5+f(x_1,\ldots,x_4)$.}

\section{Cylindricity Explained}

As a result of Proposition 2, we conclude that for all the components
$t$ (except for angular-valued ones), we have the cylindricity
condition $\partial t^{\ldots}_{\ldots}/\partial x^5=0$. Thus, the
cylindricity conditions is also explained by the geometric model. 

\section{Linear Coordinate Transformations}

In the traditional affine geometry, in addition to shifts, we can
also consider arbitrary linear coordinates transformations. In
geometric terms, we can define these transformations as continuous
automorphisms of the additive group $K_0=R^4$. We can define vectors and
tensors as continuous homomorphisms $T:K_0^p\times (K^*_0)^q\to S^1$;
in this case, e.g., standard formulas for transforming covectors
(i.e., continuous homomorphisms $g:K_0\to S^1$) can be uniquely
determined by the requirement that the value $g(a)$ be preserved under 
such a transformation, i.e., that $g'(a')=g(a)$. Similarly, the
transformation law for tensors can be determined by the condition that 
$$t'(a'_1,\ldots,a'_p,b'_1,\ldots,b'_q)=
t(a_1,\ldots,a_p,b_1,\ldots,b_1).\eqno{(1)}$$
Similarly, for $K=R^4\times S^1$, we can define a {\em $K$-linear
transformation} as follows:
\medskip

\noindent{\bf Definition 7.} {\em By a {\em $K$-linear
transformation}, we mean a continuous automorphism of the additive group of
$K$.}
\medskip

\noindent{\bf Proposition 3.} {\em Every $K$-linear transformation has
the form 
$$x^5\to \pm x^5+\sum_{i=1}^4 A_i\cdot x^i;\ \ 
x^i\to \sum_{j=1}^4 b^i_j x^j,\ \ (i\le 4).$$}
\medskip

The corresponding tensor transformations can be defined by the
condition (1). Once can see that in this case, the tensor components
are transformed just like the normal tensor components. In particular,
under the above $K$-linear transformation, a covector is transformed
as follows:
$$x_5\to \pm x_5,\ \ x_i\to \sum_{i=1}^4 c^j_i x_j-A_i\cdot x_5,$$
where $c^j_i$ is the matrix that is inverse to $b^i_j$. 

\section{General Coordinate Transformations}

\noindent{\bf Definition 8.} {\it A smooth transformation $s:K\to K$ is
{\em admissible} if and only if for each point $x\in K$, the
corresponding tangent transformation 
$$a^i\to a^i_{\rm new}=\sum_{j=1}^5 \frac{\partial s^i}{\partial x^j}_{|x}
a^i$$
is a $K$-linear transformation.}
\medskip

\noindent{\bf Proposition 4.} {\em Every admissible transformation has 
the form 
$$x^5\to \pm x^5+f(x^1,\ldots,x^4),
\ \ x^i\to f^i(x^1,\ldots,x^4).$$}
\medskip

\noindent{\it Comment.} 
We have already mentioned that functions on $K=R^4\times S^1$ are
simply functions on $R^5$ which are periodic in $x^5$ with the period 
$h$. Also, a $K$-covector $p$ can be simply viewed as a covector for which 
the fifth component $p_5$ is an integer multiple of $1/h$. Thus,,
e.g., a $K$-covector field on $K$ can be viewed as a covector field
$p(x)=(p_1(x),\ldots,p_5(x))$ on 
$R^5$ that satisfies the following two properties:
\begin{itemize}
\item[(a)] this field is periodic in $x^5$ with period $p$; 
\item[(b)] for each $x$, the value $p_5(x)$ is an integer multiple of
$1/h$. 
\end{itemize}
It is therefore reasonable to define a general coordinate
transformation of $K$ as a coordinate transformation of $R^5$ that
preserves this property, i.e., under which a covector field that
satisfies the properties (a) and (b) are transformed into a covector
field that also satisfies these properties. One can see that this
leads to the same class of general coordinate transformations. 

\section{Gauge Transformations Explained}

According to Proposition 4, every admissible transformation is a
composition of a 4D transformation and an additional gauge
transformation $x^5\to x^5+f(x^1,\ldots,x^4)$ -- exactly 
as described by Rumer. 

\section{Case of Curved Space-Time}

In modern physics, space-time is a manifold, i.e., a topological space 
$V$ which is locally diffeomorphic to $R^4$. Since our basic model is not
$R^4$, but $K=R^4\times S^1$, it is reasonable to define a {\em
$K$-manifold} as a topological space that is locally diffeomorphic to
$K$. 

From the mathematical viewpoint, $K$ is $R^5$ factorized over the 
vector $e=(0,\ldots,0,h)$: i.e., $a\sim b$ if and only if $a-b$ is an
integer multiple of $e$. Thus, a natural way to describe a
$K$-manifold is to describe a standard 5D manifold in which we have a
vector $e(x)$ in every tangent space -- i.e., a manifold with an
additional vector field. 

In this case, every tangent space is isomorphic to $K$. 
Thus, a $K$-tensor field can be defined as a mapping that maps every
point $x\in V$ into a $K$-tensor defined over the space $K$ which is
tangent at $x$. 

\section{Auxiliary Result: Why There Is No Physically Useful
Gravitational Analog of Hertz Potential}

In electromagnetism, in addition to the electromagnetic file $F_{ij}$
and the potential $A_i$ from which this filed can be obtained by
differentiation $F_{ij}=
\partial A_i/\partial x_j- \partial A_j/\partial x_i$, there is
also a useful notion of a {\em Hertz potential} $H^{ik}$ for which 
$A^i$ can be obtained by differentiation $A^i=\sum\limits_k \partial
H^{ik}/\partial x^k$. 

In gravitation, the natural analogy of potentials $A_i$ is the gravity 
tensor filed $g^{ij}$. From the purely mathematical viewpoint, it is
possible to introduce a gravitational analog of the Hertz potential:
namely, there exists a tensor field $\Pi^{ijk}$ for which 
$$g^{ij}=\sum\limits_k 
\frac{\partial \Pi^{ijk}}{\partial x^k};\eqno{(2)}$$ 
see, e.g., \cite{Palchik 1969}. 
However, in contrast to the electromagnetic case, this new potential
does not seem to have any physical applications. Why?

Our explanation is simple: while (2) is impossible in the 4D case, it
is no longer possible if we consider 5D $K$-tensor fields. 

\section*{Acknowledgments} 
The research was partially supported by NASA under 
cooperative agreement NCC5-209, by 
NSF grants EAR-0112968, EAR-0225670, and EIA-0321328, and 
by NIH grant 3T34GM008048-20S1.

The authors are thankful to 
all the participants of the special
section of the October 1997 Montreal meeting of the American
Mathematical Society, where physico-geometric aspects of this research
were presented, for valuable comments; we are especially thankful to
Prof. Abraham Ungar who organized this session, and to Yakov Eliashberg
(Stanford) for important comments.

\section*{Appendix: Proofs}

\subsection*{Proof of Proposition 1}

Let us first prove that every $K$-tensor can be described by the
desired formula. 

Indeed, let 
$t$ be a $K$-tensor. Let us first consider the restriction of $t$
to $K^p\times (R^4)^q$. 
Since locally, $K$ coincides with $R^5$, this
restriction is, locally, a multi-linear map from 
$(R^5)^p\times (R^4)^q$ to $S^1$. 
Since it is multi-linear, at 0, the value of this
map is 1. In a small vicinity of 1, we can define a unique 
angle $(1/{\rm i})\cdot \ln t$. The resulting mapping is -- locally --
a multi-linear mapping, in the traditional sense of this term, from
$(R^5)^p\times (R^4)^q$ to $R$.
Hence, in this vicinity, $\ln t={\rm i}\cdot \sum t^{\ldots}_{\ldots}
\cdot x^{i_1}\cdot \ldots$ So, for the restriction of $t$ to 
$K^p\times (R^4)^q$, we get the desired formula. 

Similarly, for $K_0\stackrel{\rm def}{=}(R^5)^m\times R^4\times
\ldots\times R^4\ldots \times e \times R^4\times \ldots\times R^4$, 
with
$r$-th term replaced by 
$e\stackrel{\rm def}{=}(0,0,0,0,h^{-1})$, we conclude that the
restriction of $t$ to $K_0$ has the form 
$$\exp\left({\rm i}\cdot 
\sum t_{i_1\ldots 
i_p}^{j_1\ldots j_{r-1}5 j_{r+1}\ldots j_q} 
\cdot x^{i_1}\cdot \ldots \cdot y^{i_p}\cdot z_{j_1}\cdot \ldots\cdot
u_{j_q}\right)$$
for some values $t_{i_1\ldots 
i_p}^{j_1\ldots j_{r-1}5 j_{r+1}\ldots j_q}$.
Since the restriction of 
$t$ to the $r$-th copy of $K^*$ is a homomorphism, this formula also
holds for elements of $(R^5)^m\times R^4\times
\ldots\times R^4\ldots \times Z\times R^4\times \ldots\times R^4$, 

Similar formulas hold for the subsets that can be obtained by replacing 
some of $K^*=R^4\times Z$ with $R^4$ and some by $Z$. Since $t$ is a
homomorphism w.r.t.~each of its variables, we can represent each
element $p=(p_1,\ldots,p_4,p_5)\in K^*$ as a sum of
$p^{(4)}=(p_1,\ldots,p_4,0)\in R^4$ and $p^{(5)}=(0,\ldots,0,p_5)\in
Z$. For each of these two vectors, we have the desired formula;
multiplying them, we get a similar formula for $p$. By using a
similar decomposition w.r.t.~other variables, we get the desired
formula for all possible inputs from $K^p\times (K^*)^q$. 

Let us now prove the desired properties of the components
$t^{\ldots}_{\ldots}$. Since $t$ is defined on $K^p\times (K^*)^q$,
replacing $x^5$ with $x^5+h$ should change the sum 
$$\sum_{i_1,i_2,\ldots,i_p,j_1,\ldots,j_q} 
t_{i_1i_2\ldots i_p}^{j_1\ldots j_q} 
\cdot x^{i_1}\cdot d^{i_2}\ldots \cdot y^{i_p}\cdot z_{j_1}\cdot \ldots\cdot
u_{j_q}$$
by an integer multiple of $2\cdot\pi$. In other words, the difference
between the new sum and old sum, i.e., 
$$h\cdot 
\sum_{5,i_2,\ldots,i_p,j_1,\ldots,j_q} 
t_{5 i_2\ldots i_p}^{j_1\ldots j_q} \cdot 
d^{i_2}\cdot \ldots \cdot y^{i_p}\cdot z_{j_1}\cdot \ldots\cdot
u_{j_q}$$
must be a multiple of
$2\cdot\pi$ for all $d^{i_2}, \ldots, y^{i_p}$. 

Let us first consider the case $p>1$. 
For $d^{i_2}=\ldots=y^{i_p}=0$, the difference is equal to 0; this
difference 
continuously depends on $d^{i_2}, \ldots, y^{i_p}$, and it is only
allowed a discrete set of values. Due to continuity, it cannot ``jump'' 
to values $2\cdot\pi\cdot k$ for $k\ne 0$, hence it is always equal to 
0. So, the above polynomial is identically 0, hence all its 
coefficients $t_{5 i_2\ldots i_p}^{j_1\ldots
j_q}$ are identically 0. 

Similarly, we can prove that $t_5^{i_1\ldots}=0$ if $i_1\ne 5$, so
$t^{5\ldots 5}_5$ is indeed the only non-zero component of
$t^{\ldots}_{\ldots}$ for which one
of the lower indices is 5. For this component, the fact that $h\cdot
t^{5\ldots 5}_5\cdot p_5\cdot \ldots\cdot p_5=2\cdot \pi\cdot k$,
where $p_5=1/h$, leads to the desired formula for $t^{5\ldots 5}_5$. 

To complete the proof, let us assume that the two sets of coefficients 
$t^{\ldots}_{\ldots}$ and $s^{\ldots}_{\ldots}$ define the same
$K$-tensor. This means that for their difference
$\delta^{\ldots}_{\ldots}$, the sum 
$$\sum_{i_1,\ldots,i_p,j_1,\ldots,j_q} 
\delta_{i_1\ldots i_p}^{j_1\ldots j_q} 
\cdot x^{i_1}\cdot \ldots \cdot y^{i_p}\cdot z_{j_1}\cdot \ldots\cdot
u_{j_q}$$
is an integer multiple of $2\cdot\pi$ for all 
$x^{i_1},\ldots,y^{i_p}\in K$ and $z_{j_1},\ldots,
u_{j_q}\in K^*$. If $p>0$, and one of the indices $j_1,\ldots,j_q$ is
different from 5, then, as above, we can conclude that the sum is
always 0, 

So, all the corresponding coefficients $\delta^{\ldots}_{\ldots}$ are
identically 0. 
The only possibly non-zero coefficient is $\delta^{5\ldots 5}$. For this
coefficient, the value $\delta^{5\ldots
5}p_5\cdot\ldots\cdot p_5$, with $p_5=1/h$, must be proportional to
$2\cdot\pi$ -- so $\delta^{5\ldots 5}\cdot (1/h)^p=2\cdot \pi\cdot
k$ for some integer $k$. 
Hence, the difference between $s^{5\ldots 5}$ and $s^{5\ldots 5}$
is indeed proportional to $2\cdot\pi\cdot h^p$. The proposition is proven. 

\subsection*{Proof of Proposition 2}

According to Proposition 1, the only possibly non-zero component of a
$K$-tensor with 5 as one of the lower indices is the component
$t^{5\ldots 5}_5$. All the values 
$\partial t_{i_1\ldots i_p}^{j_1\ldots j_q}/\partial x^5$ contain 5 as 
one of the lower indices, so the only component for which this value
can be different from 0 is the one with $p=0$ and
$i_1=\ldots=i_p=5$. For this component, $\partial t^{5\ldots
5}/\partial x^5=2\cdot\pi\cdot h^{p-1}\cdot k$. Since the $K$-tensor
field is continuously differentiable, this value cannot jump to a
different value of $k$, so this derivative is constant. Integrating
over $x^5$, we get the desired formula for the the dependence of this
component on $x^5$ -- as a linear function of $x^5$. 

\subsection*{Proof of Proposition 3}

Since $K$ locally coincides with $R^5$, its continuous automorphisms
locally coincide with continuous automorphisms $R^5\to R^5$, i.e., with 
linear transformations 
$$x^5_{\rm new}=A_5\cdot x^5+\sum_{i=1}^4 A_i\cdot x^i;\ \ 
x^i_{\rm new}=B^i\cdot x^5+\sum_{j=1}^4 b^i_j x^j.$$
If $y^5=x^5+h$ and $y^i=x^i$ for all other $i$, then $x$ and $y$
define the exact same point in $K$. Therefore, the new values $x_{\rm new}$
and $y_{\rm new}$ must also define the same point, hence $y^i_{\rm
new}=x^i_{\rm new}$ for $i=1,\ldots,4$ (hence $B^i=0$) and $y^5_{\rm
new}-x^5_{\rm new}=$ integer multiple of $h$ (hence $A_5$ is an
integer). 

Reversibility implies that $A_5^{-1}$ should also be an integer, hence 
$A_5=\pm 1.$ 

\subsection*{Proof of Proposition 4}

The condition that the tangent transformation is $K$-linear means that 
${\partial s^5}/{\partial x^5}=\pm 1$ (and due to
continuity this does not depend on the point $x$, i.e., either
it is everywhere equal to 1, or it is everywhere equal to $-1$), and 
${\partial s^i}/{\partial x^5}=0$ for $i<4$. Hence,
$s^5=\pm x_5+f(x^1,\ldots,x^4)$ and $s^i=f^i(x^1,\ldots,x^4)$ for
$i<5$. 

\subsection*{Proof of a the Statement About Hertz Potentials} 

As we have mentioned, it is possible that $\partial g^{55}/\partial
x^5\ne 0$. However, if the 
representation (2) was possible, then we would have 
$$\frac{\partial g^{55}}{\partial x^5}=
\frac{\partial^2 \Pi^{555}}{(\partial x^5)^2}+\sum_{i=1}^4 
\frac{\partial^2\Pi^{55i}}{\partial x^i\partial x^5}.$$
However, according to our general result about components of
$K$-tensors, all the terms in the right-hand side are 0s, so their sum 
cannot be equal to a non-zero value $\partial g^{55}/\partial
x^5$. 
\end{document}